\documentclass[prl,showpacs,twocolumn]{revtex4}
\usepackage{graphicx}
\usepackage{dcolumn}
\usepackage{bm}

\begin{document}

\title{Viscous cavity damping of a microlever in a simple fluid}
\author{A.~Siria$^{1,2}$, A.~Drezet$^1$, F.~Marchi$^{1,3}$, F.~Comin$^3$, S.~Huant$^1$ and J.~Chevrier$^{1}$}
\affiliation{$^{1}$ Institut N\'eel, CNRS and Universit\'e Joseph Fourier Grenoble, BP 166 38042 Grenoble Cedex 9, France\\
$^{2}$ CEA/LETI-MINATEC, 17 Avenue des Martyrs 38054 Grenoble Cedex 9, France\\
$^{3}$ ESRF, 6 rue Jules Horowitz 38043 Grenoble Cedex 9, France}

\begin{abstract}
We consider the problem of oscillation damping in air of a thermally actuated microlever as
it is gradually approached towards an infinite wall in parallel geometry. As the gap is decreased from
20 $\mu m$ down to 400 $nm$, we observe the increasing damping of the lever Brownian motion in the fluid laminar regime. This manifests itself as a linear decrease with distance of the lever quality factor accompanied by a dramatic softening of its resonance, and eventually leads to the freezing of the
CL oscillation. We are able to quantitatively explain this behavior by analytically solving the Navier-Stokes
equation with perfect slip boundary conditions. Our findings may have implications for microfluidics and micro/nano-electromechanical applications.

\end{abstract}
\pacs{47.61.Fg, 47.15.Rq, 85.85.+j, 07.79.Lh} \maketitle Micro-
and nano-scale mechanical levers are increasingly used as sensors
and actuators in a large variety of fundamental studies and
applications. Mass detection at the zeptogram scale \cite{Ekinci},
sub-attonewton force detection \cite{rugar01} and optical cooling
of microlevers \cite{cooling} are among the most spectacular
achievements of oscillating cantilevers (CLs). These realizations
mainly rely upon the extraordinary high quality factors (Q) of
oscillating CLs in vacuum and/or cryogenic temperatures where
values exceeding 100 000 are attainable. Clearly, maintaining such
performances in air or in a liquid is a very challenging issue as
oscillation damping in the surrounding fluid dramatically degrades
Q. This has been partially circumvented by using ultrasmall
self-sensing nano-electromechanical systems (NEMS, \textit{i.e.},
actuated mechanical devices made from submicron mechanical
components facing each other) operating in ambient conditions of temperature and pressure \cite{Roukes2007}.\\
\indent However, oscillating CLs are also used in viscous
environments on many occasions \cite{Charlaix1,Shih,Hansma,Aime}.
In Atomic Force Microscopy (AFM) for example, a resonant CL is
used to measure surface topography and physico-chemical properties
of various materials not only in air  but also in
liquids~\cite{Raman} for, \textit{e.g.}, visualizing dynamic biomolecular
processes at video rate~\cite{muller}. The interaction
between an AFM CL and a surrounding liquid has been used for a
distance calibration in a Casimir force measurement \cite{Cap1}
and has led very recently to the spectacular demonstration of a
repulsive Casimir force \cite{Cap3}. Therefore, the need for a
quantitative study of the CL behavior in viscous micro- and
nano-scale environments is increasing. In this letter, we report
such a quantitative study and show, down to the
submicron scale and in the demanding plane-plane geometry, how
confinement and boundary conditions at the solid-fluid interfaces conspire to change the coupling to thermal bath and how
this can freeze
out the lever oscillation.\\
\begin{figure}[hbtp]
\includegraphics[width=8cm]{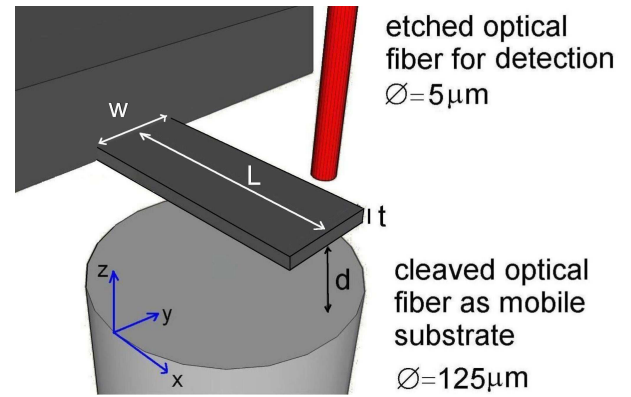}
\caption{\label{fig:setup} Scheme of the experimental setup (not
to scale). The analyzed mechanical system is a standard AFM CL. A
cleaved optical fiber (bottom) is used as mobile substrate forming
a cavity with the CL. An etched optical fiber (top) is used for
the interferometric detection of the CL Brownian motion. The main
geometrical parameters used in the text are identified, the $z$
origin is taken on the flat cleaved-fiber surface.}
\end{figure}
\indent When a CL beam vibrates in a viscous fluid, the fluid
offers resistance to the beam displacement \cite{ENS,Bhila}. If
the CL is vibrating close to a solid surface, the behavior of the
fluid and, consequently, that of the lever are modified by the
surface due to confinement. The Navier-Stokes (NS) equations give
a complete description of the fluid behavior taking into account
the particular environment under analysis. However an analytical
solution of NS equations is possible only for a restricted number
of geometries and comparison of theory with experimentally
relevant configurations is in general a complex matter or is even
lacking, especially at the deep micron and submicron scales
\cite{Green,Naik,Paul,Dorignac,Basak,Tung} where boundary
conditions at the fluid-solid interfaces are strongly
modified~\cite{Tabeling,Chan,Vinogradova,Maali}. In this work, we
focus on the dynamical behavior of a microlever close to a planar
rigid surface in the air. Provided that adapted boundary conditions are used, the NS equations can be solved
analytically for this plane-plane model geometry that mimics a
basic part of a MEMS (the counterpart of a NEMS in the micron
range) device operating in the air. This, combined with the use of
the fluctuation-dissipation theorem and of an experimental
arrangement specially designed to gain access to the intrinsic
behavior of the CL, enables us to make a quantitative comparison
between theory and experiment
in a wide range of cavity lengths down to a few hundreds of nanometers.\\
Our setup is shown schematically in Fig.~\ref{fig:setup}. Its
first specification is that the CL - a commercial thin silicon AFM CL
\cite{removed} for liquid imaging with dimensions $L\times w
\times t=107\times30\times0.18\ \mu m^3$ - is actuated by the stochastic thermal noise
only. This induces sub-Angstrom oscillations at the CL resonance
frequency ($\omega_0/(2\pi)\simeq 49.5$ kHz), thereby allowing us to consider
the fluid in the cavity in the laminar regime. Second, the planar
rigid surface facing the CL to form a parallel-plate cavity is
made of a cleaved optical fiber with a diameter of 125 $\mu m$
that is mounted over a three-axis inertial motor so as to be able
to \textit{adjust the cavity gap}. This positioning system offers
a large displacement range (8 $mm$ each axis full range) with a
good accuracy (40 $nm$ per step). Finally, the CL Brownian motion
is measured by means of a \textit{non invasive} interferometric
detection based on the use of a very thin optical fiber facing the
CL at a 2 $\mu m$ distance. This fiber has been chemically etched
so as to reduce its diameter to 5 $\mu m$. This corresponds
basically to the fiber core diameter plus a residual amount of the
optical cladding for better light guidance. The large ratio in
excess of $600$ between the areas of the cleaved and detection
fibers insures that only the cleaved one induces air confinement,
not the etched one, which is used for detection purpose only.
Therefore, no additional uncontrolled confinement and damping are
produced by the detection fiber.
\begin{figure}[hbtp]
\includegraphics[width=8cm]{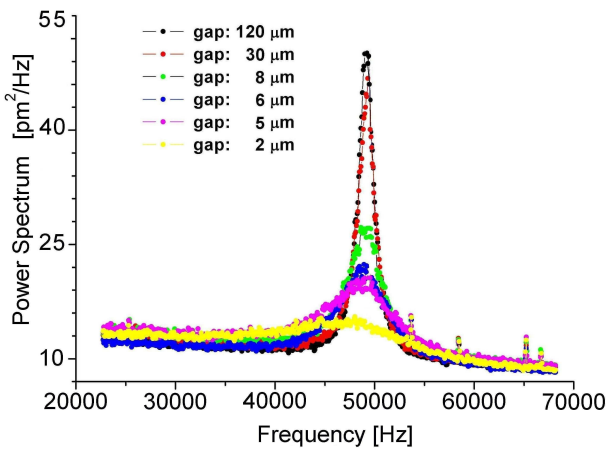}
\caption{\label{fig:brown} The experimental Brownian oscillation
power spectrum of the microlever for different cavity gaps.}
\end{figure}\\
An AFM CL vibrating in a viscous fluid may be viewed as a driven
and damped 1D harmonic oscillator whose equation of motion reads
\begin{equation}
m\ddot{z}(t)+\gamma\dot{z}(t)+kz(t)=F_{ext}, \label{eq:z1}
\end{equation} where $m$,
$z(t)$, $k$ are the CL effective mass, time-dependent position,
and stiffness, respectively, $\gamma$ is the damping factor and
$F_{ext}$ the external (\emph{i.e.}, thermal) driving force.
According to the fluctuation-dissipation theorem, the thermal
Brownian motion of the CL at temperature $T$ is accounted for by a
frequency independent force power spectrum defined as
$S_{F}(\omega)=2k_BT\gamma$ ($k_B$ is the Boltzmann constant, $\omega$ the pulsation linked to the frequency $f=\omega/(2\pi)$). Starting from Eq.~\ref{eq:z1} we
obtain the CL displacement power spectrum as $S_{z}(\omega)= S_{F}(\omega)|\chi(\omega)|^2$, where the CL transfer function $\chi(\omega)$ is given by:
\begin{equation}
\label{eq:power}
\chi(\omega)=\frac{1}{m\left(\omega_0^2-\omega^2\right)-i\gamma\omega}
\end{equation}
with $\omega_0=\sqrt{\frac{k}{m}}$ . In the
limit of small damping,
\textit{i.e.}, $\frac{\gamma}{m} \ll  \omega_0$, $S_{z}(\omega)$ has a  resonance at $\omega_0$.\\
In Fig.~\ref{fig:brown}, the experimental Brownian oscillation
power spectrum $2S_{z}(\omega)$ is presented as a function of frequency $f$ for
different cavity gaps $d$. It is clearly seen that the resonance peak
dramatically broadens and softens to lower frequencies with
decreasing gap. Within the experimental accuracy, we find that
the area under the resonance curves in Fig.~\ref{fig:brown}
remains constant and equals to the thermal energy. This shows that the CL damping increases with decreasing gap.\\
Now, we turn on to a quantitative analysis of the experiment. The
fluid responsible for the CL damping is the air confined between
the CL and the mobile fiber. The dynamic of such an incompressible
fluid is described by the NS equations
\begin{equation}
\label{eq:ns} \rho \left[ \frac{\partial \vec v}{\partial t}+\vec
v \cdot\nabla \vec v\right] = \eta \nabla ^2 \vec v -\nabla p,
\end{equation}
where $\vec v$ is the fluid velocity, $\rho$ its density, $\eta$
its dynamical viscosity and $p$ the gas pressure. In the laminar
regime, \emph{i.e.}, in the limit of small Reynolds numbers, Eq.~\ref{eq:ns} simplifies to $\eta \nabla ^2 \vec v
\simeq \nabla p$. In order to solve the NS equations, one needs to
know the specific boundary conditions existing at the
fluid-solid interfaces (for simplicity we assume the cavity plates to be infinitely extended). While for macroscopic hydrodynamic
applications one usually accepts that fluids do not slip against solid
walls, this is generally not true for microfluidic problems
involving MEMS or NEMS \cite{Tabeling}. A critical parameter in
this respect is the Knudsen number~\cite{Tabeling}
$K_n=\bar{\lambda}/d$ which depends on the gas mean free path
$\bar{\lambda}$. For air at ambient conditions
$\bar{\lambda}\simeq 60$ nm which leads here to $K_n\sim
0.001-0.06$. In this range of $K_n$ values, it is already known
that fluid-slip can
occur over a solid interface~\cite{Tabeling,Chan,Vinogradova}. In
particular, partial fluid-slip has been recently observed in the
plane-sphere geometry in air using an AFM in dynamic mode
\cite{Maali}. However, none of the previous works investigated the
regime of Brownian oscillations with typical CL amplitudes $\delta
z\sim 0.05$ nm much smaller than $\bar{\lambda}$ (\emph{i.e.}, $\bar{\lambda}/\delta z\sim
10^3$). In such a regime, boundary conditions are expected to be
even more strongly modified~\cite{Tabeling} compared with the macroscopic regime although it is not yet known how much they are modified. Here, we make the hypothesis of perfect slip, historically anticipated by Navier~\cite{Navier}, for which friction along the solid interface is prohibited, and show that we can obtain a consistent quantitative description of our experimental data. This hypothesis results in a velocity gradient along the $z$ direction
which leads to a Stokes friction coefficient:
\begin{equation}
\label{eq:gam}
 \gamma= \frac{2 \eta  A}{d},
\end{equation}
where $A=wL$ is the cantilever surface. The usual no-slip condition at the fluid solid interface would predict $\gamma\simeq\eta wL^3/d^3$, in clear disagreement with the experiment. As a direct consequence, Eq.~4 leads to a much smaller decay of the friction force with $d$ than predicted usually.  \\
Quantitative information on the damping factor is obtained from
the analysis of the CL quality factor. Both quantities are
linked together by the relation $Q= \frac{k}{\omega_0\gamma}$,
which becomes for small gaps:
\begin{equation}
\label{eq:Q}
 Q= \frac{k}{2\omega_0\eta A}d.
\end{equation}
Eq.~\ref{eq:Q} predicts a linear dependence of $Q$ with $d$ that can be compared with experiment.\\
\begin{figure}[hbtp]
\includegraphics[width=8cm]{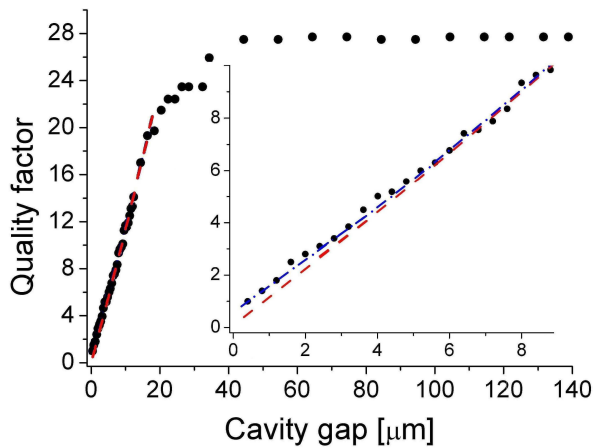}
\caption{\label{fig:quality} The quality factor as a function of
the cavity gap. The black dots are the experimental points.  The
insert depicts a zoom of the small gap range. In both cases, the
red curves exhibit the theoretical prediction based on
Eq.~\ref{eq:Q}, \textit{i.e.}, the prediction of the NS equation
in the perfect parallel plate geometry. The blue dash-dotted line
in the insert shows the prediction of the NS model taking into
account a residual angular misalignment of the cavity as discussed
in the text.}
\end{figure}
Fig.~\ref{fig:quality} depicts the quality factor as function of the cavity gap. Two different regimes can be distinguished. For large gaps above 40 $\mu m$, the quality factor remains constant. This is the unconfined fluid regime where no additional damping can take place with decreasing gap. For smaller gaps however, the quality factor tends to decrease with a decreasing gap. We will focus below on the small gap regime where the hypothesis of infinite planes is physically justified. Since the AFM CL has a surface 10 times smaller than the substrate fiber, the gap limit for the hypothesis of infinite planes to remain physically sound can be estimated by taking the apparent CL surface as reference, \textit{ i.e.} $d_{\textrm{lim.}}\approx15\mu m$.\\
As shown in the inset of Fig.~\ref{fig:quality} the experimental results and the theoretical prediction of Eq.~\ref{eq:Q}, with no adjustable parameter~\cite{measure}, coincide to within $5\%$ for gaps larger than 5 $\mu m$ but for smaller separations, the agreement worsens to reach $100\%$ at the smallest gap, 400 $nm$. We interpret this difference with a residual small angular misalignment of the two facing parallel plates.
\begin{figure}[hbtp]
\includegraphics[width=7cm]{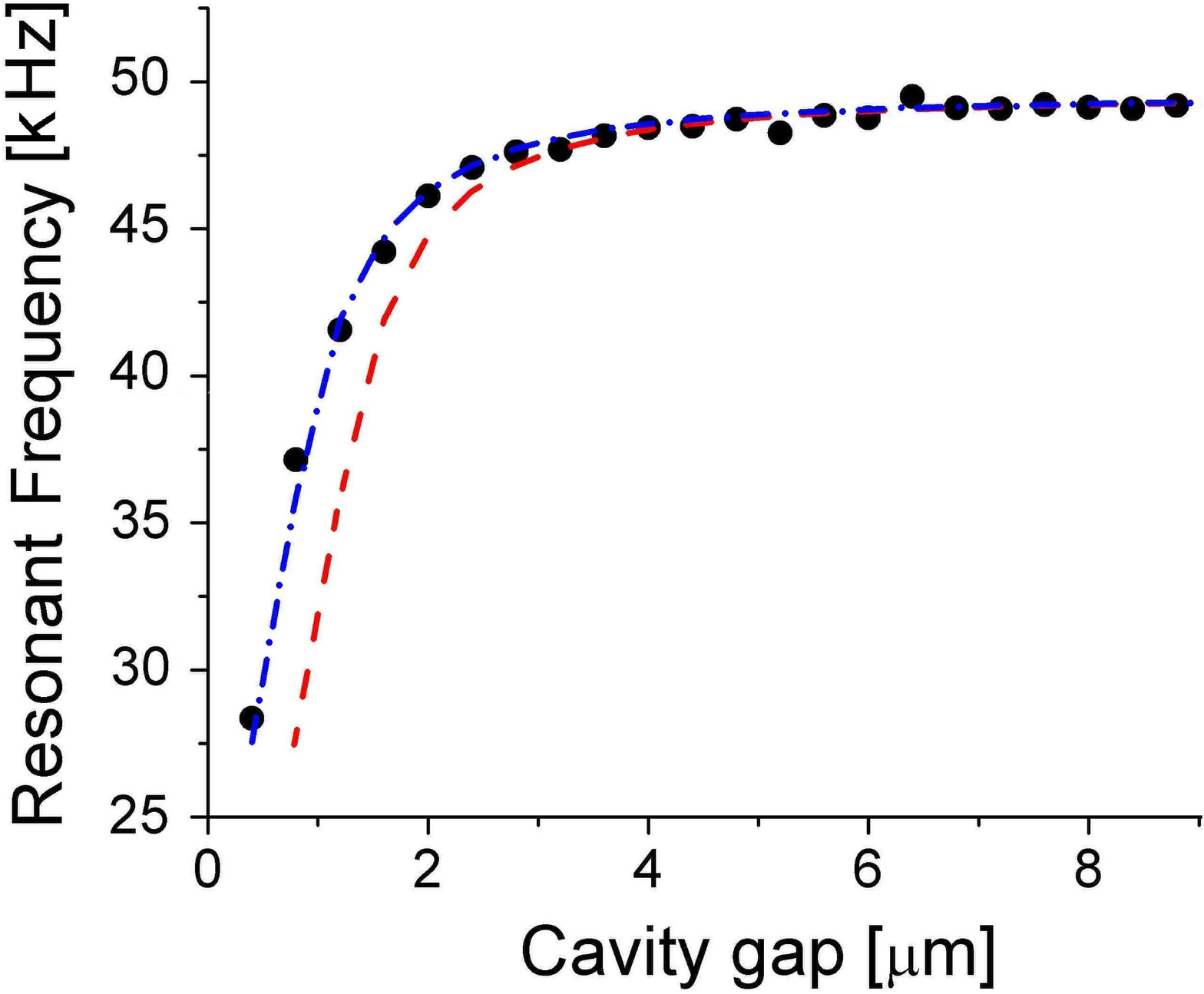}\\
\caption{\label{fig:angle} The resonance frequency as function of the cavity gap in the small gap regime. Like in Fig.\ref{fig:quality}, the red dashed, respectively blue dash-dotted curves is the prediction of the NS model for the perfectly aligned, respectively slightly misaligned, cavity.}
\end{figure}
For small misalignment, the problem can be treated within an approximation similar to the Proximity Force Approximation (PFA) used, for instance, in the Casimir force formulation in the sphere-plane geometry \cite{PFA}. In this approximation, the corrected damping factor becomes:
\begin{equation}
\gamma=\int_0^L\int_0^w 2\eta \frac{dx \cdot dy}{d_0+x\tan{\alpha}+y\tan{\beta}}
\end{equation}
where $d_0$ is the shortest distance from the inclined CL to the substrate, $\alpha$ and $\beta$ are the lateral tilt angles of the CL with the mobile surface in the $x$ and $y$ directions, respectively. The angle values that permit to reproduce the evolution of the disagreement between experiment and theory, as presented in Fig.~\ref{fig:quality}, are $\alpha\approx\beta\approx10\ mrad$. Considering these misalignment angles, the good agreement between theory and experiment can now be extended down to the smallest gap range that we have measured as can be seen in Fig.~\ref{fig:quality}. Over the entire range $400\ nm-15\ \mu m$, the remaining disagreement is $\approx5\%$ only \cite{pachyderme}.\\
We now discuss the frequency softening of the CL oscillation, the other salient experimental fact revealed by Fig.~\ref{fig:brown}. In the limit of large damping, \textit{i.e.} the approximation $\frac{\gamma}{m} \ll \omega_0$ no longer holds, the power spectrum of Eq.~\ref{eq:power} has a down-shifted resonance pulsation $\omega'$ given by:
\begin{equation}
\label{eq:newres}
\omega'=\sqrt{\omega_0^2-\frac{1}{2}\left(\frac{\gamma}{m}\right)^2}
\end{equation}
Fig.~\ref{fig:angle} shows that the resonance frequency shift
$(\omega'-\omega_0)/(2\pi)$ can be extremely large. One step further below $400 nm$, a complete freezing of the CL
oscillation would have been observed. This was precluded by the residual angular misalignment discussed above. Taking into account for data
analysis of the misalignment obtained from Fig.~\ref{fig:quality}, we can quantitatively model the measurements in Fig.~\ref{fig:angle}
without any adjustable parameter whatsoever. Therefore, beside the CL width that governs the $Q$ factor behavior at
large scale (Fig.~\ref{fig:quality}), another shorter characteristic length is here emphasized in the submicron range, i.e. $d_{\textrm{crit.}}=\frac{\sqrt{2}\eta}{m}\frac{A}{\omega_0}$ (around 500 nm in our case), which is the gap width cancelling the resonance frequency in Eq.~\ref{eq:newres}. This characteristic length is determined by the CL dynamics and the fluid viscosity.\\
In conclusion, we have presented high sensitivity measurements of
the damping of a thermally driven CL in a simple fluid confined in
a microcavity formed by this CL facing an infinite wall. As the
cavity length decreases, the fluid confinement induces a dramatic
damping of the CL Brownian motion which can lead to its complete
freezing at small gaps. A consequence of our work is that micro- or nano-oscillators can
either present high Q factors or be overdamped systems depending
on their actual geometry, resonance frequency, oscillator
substrate gap and, of course, ambient viscosity. These findings may impact the design of modern NEMS and microfluidic devices since the $1/d$ dependence strongly reduces dissipation even for separations $d$ as large as thousands of mean free path $\bar{\lambda}$ (see Fig.~3). This $1/d$ behavior can be furthermore described
by solving the Navier-Stokes equation with perfect solid-fluid
slip boundary conditions. The agreement between experiment and our model is found over a broad
range of cavity lengths, including the submicron range. Interesting extensions of the present work include the study of parameters affecting boundary conditions, such as external actuation of CLs (to obtain large oscillation amplitude)~\cite{newraman}, nanostructuration~\cite{Roukes2007}, and surface chemical properties~\cite{Aime}.\\
We are grateful to Giovanni Ghigliotti for helpful discussions. Our thin etched optical fiber has been prepared
by Jean-Francois Motte. This research was partly supported by a ``Carnot-NEMS'' collaborative grant between CEA-LETI and Institut N\'eel.


\begin{thebibliography}{}
\bibitem{removed}
The CL tip has been removed by focused-ion-beam milling.
\bibitem{measure}
The CL dimensions have been measured by electron microscopy. The air viscosity was taken as $\eta\simeq 1.8\times 10^{-5} kg/ms$. The CL stiffness $k=0.0396N/m$ has been experimentally determined from Fig. \ref{fig:brown} using the energy equipartition theorem.
\bibitem{pachyderme}
Following Ref.~\cite{Aime}, the ultra small CL oscillation precludes any mass addition to the lever, in contrast with Ref. \cite{Naik} where this effect is dominant for actuated macroscopic levers.
\end{thebibliography}
\end{document}